\documentstyle[12pt,epsf]{article}

\advance \topmargin by -\headheight
\advance \topmargin by -\headsep

\evensidemargin \oddsidemargin
\newcommand{\beq}{\begin{equation}}
\newcommand{\eeq}{\end{equation}}
\newcommand{\rc}{\nonumber\\}
\newcommand{\bear}{\begin{eqnarray}}
\newcommand{\eear}{\end{eqnarray}}

  \def\tr{{\hbox{\rm Tr}}}

\def\ie{{\em i.e.}}
\def\ie{\hbox{\it i.e.}}

\def\CC{{\mathchoice
{\rm C\mkern-8mu\vrule height1.45ex depth-.05ex
width.05em\mkern9mu\kern-.05em}
{\rm C\mkern-8mu\vrule height1.45ex depth-.05ex
width.05em\mkern9mu\kern-.05em}
{\rm C\mkern-8mu\vrule height1ex depth-.07ex
width.035em\mkern9mu\kern-.035em}
{\rm C\mkern-8mu\vrule height.65ex depth-.1ex
width.025em\mkern8mu\kern-.025em}}}

\def\RR{{\rm I\kern-1.6pt {\rm R}}}

\def\ZZ{{\rm Z}\kern-3.8pt {\rm Z} \kern2pt}
\def\IB{\relax{\rm I\kern-.18em B}}
\def\ID{\relax{\rm I\kern-.18em D}}
\def\II{\relax{\rm I\kern-.18em I}}
\def\IP{\relax{\rm I\kern-.18em P}}

\def\prl{Phys. Rev. Lett.}
\def\pr{Phys. Rev.}

\def\mpl{Mod. Phys. Lett.}

\def\jhep{J. High Energy Phys.}

\begin{document}

\begin{titlepage}

\begin{center} \Large \bf Mesons and Higgs branch in defect theories

\end{center}

\vskip 0.3truein
\begin{center}
Daniel Are\'an${}^{\,*}$\footnote{arean@fpaxp1.usc.es}, 
Alfonso V. Ramallo${}^{\,*}$\footnote{alfonso@fpaxp1.usc.es} and
Diego Rodr\'\i guez-G\'omez${}^{\,\dagger}
$\footnote{Diego.Rodriguez.Gomez@cern.ch}

\vspace{0.3in}

${}^{\,*}$Departamento de F\'\i sica de Part\'\i culas, Universidade de
Santiago de Compostela \\and\\
Instituto Galego de F\'\i sica de Altas Enerx\'\i as (IGFAE)\\
E-15782 Santiago de Compostela, Spain
\vspace{0.3in}

${}^{\,\dagger}$Departamento de F\'\i sica, Universidad de Oviedo\\
Avda. Calvo Sotelo 18, 33007 Oviedo, Spain

\end{center}
\vskip.5
truein

\begin{center}
\bf ABSTRACT
\end{center}
We consider the defect theory obtained by intersecting D3- and D5-branes
along two common spatial directions. We work in the approximation in which
the D5-brane is a probe in the $AdS_5\times S^5$ background. By adding
worldvolume flux to the D5-brane and choosing an appropriate embedding
of the probe in  $AdS_5\times S^5$, one gets a supersymmetric
configuration in which some of the
D3-branes recombine with the D5-brane. We check this fact by showing that
the D5-brane can be regarded as a system of polarized D3-branes. On
the field theory side this corresponds to  the Higgs branch of the
defect theory, where some of the fundamental hypermultiplet fields
living on the intersection acquire a vacuum expectation value. We study
the spectrum of mesonic bound states of the defect theory in this Higgs
branch and show that it is continuous and gapless.

\vskip5.6truecm
\leftline{US-FT-4/06}
\leftline{FFUOV-09-10}
\leftline{hep-th/0609010 \hfill September  2006}
\smallskip
\end{titlepage}
\setcounter{footnote}{0}

\section{Introduction}	
One of the most exciting developments in the study of the gauge/gravity
correspondence \cite{jm,MAGOO} is the extension of this duality to include
open string degrees of freedom, which corresponds, on the gauge theory
side, to adding matter fields in the fundamental representation of the
gauge group.  The standard way to perform this extension is by including
D-branes (flavor branes) in the supergravity background \cite{KR,KKW}. If
the number of extra D-branes is small,  one can neglect their backreaction
on the background and treat them as probes whose fluctuation modes are
identified with the mesonic bound states of the theory with flavor (for
a review see \cite{MPLAreview} and references therein).

The best studied flavor brane system is the one corresponding to the
D3-D7 intersection which, in the decoupling limit, is equivalent to a
D7-brane embedded in the $AdS_5\times S^5$ background such that, in the
UV, the induced worldvolume metric is of the form $AdS_5\times S^3$. 
In the probe approximation the fluctuation modes of this
system can be integrated analytically and the corresponding meson spectra
can be obtained exactly \cite{KMMW}. In a more recent progress
\cite{Higgs} the meson spectrum of this system in a mixed Coulomb-Higgs
branch has been obtained. In this branch some  fundamental
hypermultiplet fields have non-vanishing vacuum expectation values. In
the dual supergravity description the Higgs branch is described by
instanton configurations of the worldvolume gauge field. This instantonic
gauge field lives on the directions of the D7-brane worldvolume which are
orthogonal to the gauge theory directions. The meson spectra for the
case of two flavors has been computed in ref. \cite{Higgs} by using the
non-abelian Dirac-Born-Infeld action with an $SU(2)$ instanton. The
corresponding mass levels depend on the size of the instanton and display
an spectral flow phenomenon.

In this paper we will perform a similar analysis for the supersymmetric
intersection of D3- and D5-branes, according to the array:
\beq
\begin{array}{ccccccccccl}
 &1&2&3& 4& 5&6 &7&8&9 & \nonumber \\
D3: & \times &\times &\times &\_ &\_ & \_&\_ &\_ &\_ &     \nonumber \\
D5: &\times&\times&\_&\times&\times&\times&\_&\_&\_ &
\end{array}
\label{D3D5intersection}
\eeq
Notice that the D5-brane is of codimension one along the gauge theory
directions of the D3-brane worldvolume. Actually, this D3-D5 system is
dual \cite{KR} to a defect theory in which ${\cal N}=4$, $d=4$ super
Yang-Mills theory in the bulk is coupled to ${\cal N}=4$, $d=3$
fundamental hypermultiplets localized at the defect \cite{WFO, EGK},
which is located at a fixed value of the coordinate 3 in the array 
(\ref{D3D5intersection}). These
hypermultiplets  arise from the  3-5 open strings and their mass is
proportional to the separation of the two stacks of branes in the 789
directions of the array (\ref{D3D5intersection}). When this distance is
zero, the induced metric on the worldvolume of the D5-brane is
$AdS_4\times S^2$, while for non-zero distance we introduce a mass scale
which breaks conformal invariance and, as a consequence, the induced
metric is $AdS_4\times S^2$ only asymptotically in the UV. The meson
spectra in this latter case can be computed analytically \cite{open} (see
also \cite{MT}).

By switching on a
worldvolume magnetic field along the $S^2$,  one can still have a
supersymmetric configuration if the D5-brane is appropriately bent along
the 3 direction \cite{ST}, which corresponds to a different 
$AdS_4\times S^2\subset AdS_5\times S^5$ embedding. 
This worldvolume field induces D3-brane charge
to the D5-brane probe. Actually, we will show that, in this case, the
D5-brane probe can be regarded as a bound state of polarized D3-branes.
Moreover, by looking directly at the action of ref. \cite{WFO} for the
defect field theory, we will verify that our configuration corresponds to
a situation in which some of the D3-branes end on a D5-brane and recombine
with it. This recombination realizes the Higgs branch of the theory, in
which some components of the fundamental hypermultiplets acquire a
non-vanishing vacuum expectation value. Finally, we will look at the
fluctuations of the probe around the static configuration. We will show
that, contrary to what happens to the D3-D7 system, there is no discrete
spectrum for the meson masses in this D3-D5 intersection with flux. The
reason behind this result is the fact that our fluctuations are not
localized at the defect and, instead, they spread over the direction 3 of
the D3-brane worldvolume.  

This paper is organized as follows. In section \ref{fluxD3D5} we
introduce the D3-D5 configuration with flux we are interested in. In
section \ref{dielectric} we show that the D5-brane in this
configuration admits a microscopic interpretation as bound state of
D3-branes. The field theory analysis and the relation to the Higgs branch
of the defect theory are the subjects of section \ref{fieldtheory}. The
fluctuations are studied in section
\ref{fluctuations}. Finally, we end with some concluding remarks in
section \ref{conclusions}.

\section{The D3-D5 intersection with flux}	
\label{fluxD3D5}
The near-horizon limit of the metric generated by a stack of $N$ parallel
D3-branes is $AdS_5\times S^5$, which we will write as:
\beq
ds^2\,=\,{r^2\over R^2}\,\,dx_{1,3}^2\,+\,{R^2\over r^2}\,
d\vec y\cdot d\vec y\,\,,
\label{backgroundmetric}
\eeq
where $\vec y\,=\,(\,y^1,\cdots ,y^6)\,\,$ are the six coordinates
orthogonal to the worldvolume of the D3's, 
$r^2\,=\,\vec y\cdot
\vec y$  and the radius $R$ is given by:
\beq
R^4\,=\,4\pi\, g_s\,N\,(\alpha')^2\,\,.
\eeq
In eq. (\ref{backgroundmetric}) $dx_{1,3}^2$ represents the
$(3+1)$-dimensional Minkowski metric for the coordinates 
$x^0,\cdots, x^3$. Moreover, the D3-brane background is endowed with a
Ramond-Ramond five-form $F^{(5)}$, whose potential will be denoted by
$C^{(4)}$. The component of $C^{(4)}$ along the Minkowski coordinates
is given by:
\beq
\Big[\,C^{(4)}\,\Big]_{x^0\cdots x^3}\,=\,
\Biggl[\,{r^2\over R^2}\,\Biggr]^{2}\,\,.
\label{CRR}
\eeq
For convenience, let us split the six $\vec y$ coordinates in two sets of
three elements, according to the D3-D5 intersection represented by
the array (\ref{D3D5intersection}). The coordinates $(\,y^1,y^2 ,y^3)$ are
those which are parallel to the D5-brane worldvolume in
(\ref{D3D5intersection}). We will write their contribution to the
line element in (\ref{backgroundmetric}) in spherical coordinates as 
$(dy^1)^2+(dy^2)^2+(dy^3)^2\,=\,d\rho^2\,+\,\rho^2 d\Omega_2^2$, where 
$d\Omega_2^2$ is the line element of a unit two-sphere. Moreover, let us
denote by $\vec z\,=\,(\,z^1,z^2 ,z^3)\,=\,(\,y^4,y^5 ,y^6)$ the
coordinates transverse to  both  the D3- and D5-branes. Clearly, 
$r^2\,=\,\rho^2+\vec z^{\,\,2}$ and the $AdS_5\times S^5$ metric
(\ref{backgroundmetric}) can be
written as:
\beq
ds^2\,=\,{\rho^2+\vec z^{\,\,2}\over R^2}\,\,dx_{1,3}^2\,+\,
{R^2\over \rho^2+\vec z^{\,\,2}}\,\bigg(\,
d\rho^2\,+\,\rho^2 d\Omega_2^2\,+\,d\vec z\cdot d\vec z\,\bigg)\,\,.
\label{polarbackmetric}
\eeq

The action of a D5-brane probe in the $AdS_5\times S^5$ background is
given by the sum of the Born-Infeld and Wess-Zumino terms, namely:
\beq
S_{D5}\,=\,-\,T_{5}\,\int d^6\xi\,\sqrt{-\det (g+F)}\,+\,
T_{5}\,\int d^6\xi \,\,\,P\big[\,C^{(4)}\,\big]\wedge F\,\,,
\label{DBI-D5}
\eeq
where $T_{5}$ is the tension of the D5-brane, given by 
$T_{5}^{-1}\,=\,(2\pi)^5\,(\alpha')^3\,g_s$, 
$g$ is the pullback of the metric (\ref{polarbackmetric}), $F$ is
the strength of the abelian worldvolume gauge field and $\xi^a$
$(a=0,\cdots, 5)$ are a set of worldvolume coordinates. In what follows
we will use $x^0, x^1, x^2$ and the radial ($\rho$) and angular
coordinates of eq. (\ref{polarbackmetric}) as our set of worldvolume
coordinates. The embedding  of the D5-brane probe is then specified by 
the values of $x^3$ and $\vec z$ as functions of the $\xi^a$'s. We will
consider static embeddings in which $|\vec z|$ is a fixed constant, namely
$|\vec z|=L$. The simplest of such embeddings is the one in which the
coordinate $x^3$ is also a constant and the worldvolume gauge field $F$
vanishes. This configuration was proposed in ref. \cite{KR}, and studied
extensively in ref.
\cite{WFO}, as a prototype of a defect theory. In this case the defect is
a flat wall determined by the condition $x^3={\rm constant}$ and the
induced worldvolume metric is, for large $\rho$, of the form $AdS_4\times
S^2$. The corresponding dual field theory is ${\cal N}=4$, $d=4$ super
Yang-Mills theory coupled to
${\cal N}=4$, $d=3$ hypermultiplets localized at the defect which are in
the fundamental representation. Let us generalize this configuration of
the D5-brane probe by switching on a magnetic field $F$ along the
two-sphere of its worldvolume. To be precise, let us assume that $F$ is
given by:
\beq
F\,=\,q {\rm Vol}\,(S^2)\,\equiv\,{\cal F}\,,
\label{wvflux}
\eeq
where $q$ is a constant and ${\rm Vol}\,(S^2)$ is the volume form of the 
worldvolume two-sphere.  To understand the implications of having a
magnetic flux across the worldvolume $S^2$, let us look at the form of
the Wess-Zumino term in the action (\ref{DBI-D5}), namely:
\beq
S_{WZ}\,\sim\,\int_{S^2}\,F\,\int P[\,C^{(4)}\,]\,\sim q\,x'\,\,,
\label{WZbending}
\eeq
where  $x\equiv x^3$ and the prime denotes the derivative  with respect
to the radial coordinate $\rho$. It is clear by inspecting the right-hand
side of eq. (\ref{WZbending}) that the worldvolume flux acts as a source
of  a non-trivial dependence of $x$ on the coordinate $\rho$. Actually,
assuming that $x$ only depends on $\rho$, the action (\ref{DBI-D5}) of the
probe takes the form:
\beq
S_{D5}=-4\pi\,T_{5}\,\int\,d^3x\,d\rho\Bigg[\,
\rho^2\,
\sqrt{1\,+\,{(\rho^2+L^2)^2\over R^4}\,x'^{\,2}}\,\,
\sqrt{1\,+\,{(\rho^2+L^2)^2\over R^4}\,
{q^2\over \rho^4}}\,\,-\,\,
{(\rho^2+L^2)^2\over R^4}\,q\,x'\,\Bigg]\,\,,
\label{effe-action}
\eeq
where we have assumed that $\vec z$ is constant ($|\,\vec z\,|=L$) and we
have integrated over the coordinates of the two-sphere. The
Euler-Lagrange equation for $x(\rho)$ derived from (\ref{effe-action}) is
quite involved. However, there is a simple first-order equation for
$x(\rho)$ which solves this equation \cite{ST} , namely:
\beq
x'(\rho)\,=\,{q\over \rho^2}\,\,.
\label{first-order}
\eeq
Actually, the first-order equation (\ref{first-order}) is a BPS equation
required by supersymmetry, as can be verified by checking the kappa
symmetry of the embedding \cite{ST}. The integration of eq.
(\ref{first-order}) is straightforward:
\beq
x(\rho)\,=\,x_0\,-\,{q\over \rho}\,\,,
\label{bending}
\eeq
where $x_0$ is a constant. The dependence on $\rho$ of the right-hand
side of eq. (\ref{bending}) represents the bending of the D5-brane
profile required by supersymmetry when there is a non-vanishing flux of
the worldvolume gauge field. Notice also that now the probe is located at
a fixed value of $x$ only at the asymptotic value $\rho\to\infty$, whereas
when $\rho$ varies the $D5$-brane fills one-half on the worldvolume of
the D3-brane (\ie\ $x^3\le x_0$ for $q>0$). 

It is also interesting to study the modifications of the induced metric
introduced by the bending. Actually, when $q\not=0$  this induced metric
takes the form:
\beq
{\cal G}_{ab}\,d\xi^a d\xi^b\,=\,
{\rho^2+L^2\over R^2}\,dx^2_{1,2}\,+\,
{R^2\over \rho^2+L^2}\,\Bigg[\,
\bigg(1+{q^2\over R^4}\,{(\rho^2+L^2)^2\over \rho^4}\,\bigg)
d\rho^2\,+\,\rho^2\,
d\Omega_2^2\,\Bigg]\,\,.
\label{ind-met-flux}
\eeq
It can be readily verified from (\ref{ind-met-flux}) that the UV metric
at $\rho\to\infty$ (or, equivalently, when the D3- and D5-branes are at
zero distance $L$) takes the form:
\beq
AdS_4(R_{eff})\times S^2 (R)\,\,,
\label{UVind-met-flux}
\eeq
where the radius of the $AdS_4$ changes from its fluxless value $R$ to 
$R_{eff}$, with the latter given by:
\beq
R_{eff}\,=\,\bigg(\,1\,+\,{q^2\over R^4}\,\bigg)^{{1\over 2}}\,\,R\,\,.
\label{Reff}
\eeq
Notice that the radius of the $S^2$ is not affected by the flux, as is
clear from (\ref{ind-met-flux}).

One can understand the appearance of this UV metric as follows.
Let us suppose that we have an $AdS_5$ metric of the form:
\beq
ds^2_{AdS_5}\,=\,{\rho^2\over R^2}\,\,dx^2_{1,3}\,+\,
{R^2\over \rho^2}\,\,d\rho^2\,\,.
\label{AdSmetric}
\eeq
Let us now change variables from $(\rho, x^3)$ to new
coordinates $(\varrho, \eta)$, as follows:
\beq
x^3\,=\,\bar x\,-\,{\tanh\eta\over \varrho}\,\,,\qquad\qquad
\rho\,=\,R^2\varrho\cosh\eta\,\,,
\label{changevariables}
\eeq
where $\bar x$ is a constant. 
It can be easily seen that the $AdS_5$ metric (\ref{AdSmetric})
in the new variables takes the form:
\beq
ds^2_{AdS_5}\,=\,R^2\,(\,\cosh^2\eta\,ds^2_{AdS_4}\,+\,d\eta^2\,)\,\,,
\label{foliation}
\eeq
where $ds^2_{AdS_4}$ is the metric of $AdS_4$ with unit radius, given by:
\beq
ds^2_{AdS_4}\,=\,\varrho^2\,dx^2_{1,2}\,+\,{d\varrho^2\over
\varrho^2}\,\,.
\eeq
Eq. (\ref{foliation}) shows clearly the foliation of $AdS_5$ by 
$AdS_4$ slices with $\eta={\rm constant}$. The effective radius of the 
$AdS_4$ slice depends on the value of $\eta$ as follows:
\beq
R_{eff}\,=\,R\cosh\eta\,\,.
\label{sliceradius}
\eeq
It can be straightforwardly checked by using the change of variables
(\ref{changevariables}) with $\bar x=x_0$ that our embedding
(\ref{bending}) corresponds to one of these $AdS_4$ slices with a
constant  value of $\eta$ given by:
\beq
\eta\,=\,\eta_q\,=\,\sinh^{-1}\,\Big({q\over R^2}\Big)\,\,.
\eeq
Moreover, one can verify that the  $AdS_4$ radius $R_{eff}$ of eq. 
(\ref{sliceradius}) reduces to the expression given in (\ref{Reff}) when
$\eta=\eta_q$.

The worldvolume gauge field (\ref{wvflux}) is constrained by a flux
quantization condition \cite{Flux} which, with our notations, reads:
\beq
\int_{S^2}\,F\,=\,{2\pi k\over T_f}\,\,,\qquad
k\in \ZZ\,\,,\qquad T_f\,=\,{1\over 2\pi\alpha'}\,\,.
\label{fluxquantization}
\eeq
It is now immediate to conclude that the condition
(\ref{fluxquantization}) restricts the constant $q$ to be of the form:
\beq
q\,=\,k\pi\alpha'\,\,,
\label{q-k}
\eeq
where $k$ is an integer.

\section{Dielectric interpretation}	
\label{dielectric}

The presence of a worldvolume flux as in (\ref{wvflux}) induces, through
the Wess-Zumino term of the action (\ref{DBI-D5}), a D3-brane charge,
proportional to $\int_{S^2}\,F$, on the D5-brane.
For this reason it
is not surprising that the D5-brane configuration of section
\ref{fluxD3D5} admits a microscopical description in terms of a bound
state of coincident D3-branes. Actually,  the integer
$k$ of the quantization condition (\ref{fluxquantization}) has the
interpretation of the number of D3-branes that build up the D5-brane. The
dynamics of a stack of coincident D3-branes is determined by the Myers
dielectric action \cite{dielectric}, which is the sum of a Born-Infeld and
a Wess-Zumino part:
\beq
S_{D3}\,=\,S_{BI}^{D3}\,+\,S_{WZ}^{D3}\,\,.
\eeq
For the background we are considering the Born-Infeld action is:
\beq
S_{BI}^{D3}\,=\, -T_3\,\int\,\,d^4\xi\,\, {\rm Str}\,\Bigg[\,
\sqrt{-\det\bigg[ P[G+G(Q^{-1}-\delta)G]_{ab}\,\bigg]}\,
\sqrt{\det Q}\,\Bigg]\,\,,
\label{dielectricBI}
\eeq
where we have set the worldvolume gauge field to zero. In eq. 
(\ref{dielectricBI}) $T_3$ is the tension of the D3-brane, given by
$T_3^{-1}\,=\,(2\pi)^3\, (\alpha')^2\,g_s$,  and $G$ is the background
metric (\ref{polarbackmetric}). In this dielectric picture the D3-brane
has non-commutative transverse scalars represented by matrices. In eq. 
(\ref{dielectricBI}) ${\rm Str}(\cdots)$ represents the symmetrized trace
and $Q$ is a matrix which depends on the commutator of the transverse
scalars (see below). The Wess-Zumino term for the D3-brane in the
$AdS_5\times S^5$ background under consideration is:
\beq
S_{WZ}^{D3}\,=\,
T_{3}\,\int d^4\xi \,\,\,{\rm Str}\,
\bigg[\,P\big[\,C^{(4)}\,\big]\,\bigg]\,\,.
\label{dielectricWZ}
\eeq
Let us now choose $x^0,x^1,x^2$ and $\rho$ as our set of worldvolume
coordinates of the D3-branes. Moreover, we shall introduce new
coordinates 
$Y^I(I=1,2,3)$ for the two-sphere of the metric (\ref{polarbackmetric}).
These new coordinates satisfy $\sum_I\,Y^I\,Y^I\,=\,1$ and
the line element $d\Omega_2^2$ is given by:
\beq
d\Omega_2^2\,=\,\sum_I\,dY^I\,dY^I\,\,,
\qquad\qquad
\sum_I\,Y^I\,Y^I\,=\,1\,\,.
\eeq
We will assume that the $Y^I$'s are the only non-commutative scalars.
They will be represented by $k\times k$ matrices.  In this case the 
matrix $Q$ appearing in (\ref{dielectricBI}) is given by:
\beq
Q_J^I\,=\,\delta_{J}^{I}\,+\,{i\over 2\pi\alpha'}\,
[Y^I,Y^K]\,G_{KJ}\,\,.
\eeq
Actually, we shall adopt the  ansatz in which the  $Y^I$'s are constant
and given by:
\beq
Y^I\,=\,{J^I\over \sqrt{C_2(k)}}\,\,,
\label{Yansatz}
\eeq
where the $k\times k$ matrices $J^I$ correspond to the $k$-dimensional
irreducible representation of the $SU(2)$ algebra:
\beq
[J^I,J^J]\,=\,2i\epsilon_{IJK}\,J^K\,\,,
\label{Jcommutator}
\eeq
and $C_2(k)$ is the quadratic Casimir of the $k$-dimensional irreducible
representation of  $SU(2)$ ($C_2(k)=k^2-1$). Therefore, the $Y^I$ scalars
parametrize a fuzzy two-sphere. Moreover, let us assume that we consider
embeddings in which the scalars $\vec z$ and $x^3$ are commutative and
such that $|\vec z|=L$ and $x^3=x(\rho)$ (a unit $k\times k$ matrix is
implicit). With these conditions, as the metric (\ref{polarbackmetric})
does not mix the directions of the two-sphere with the other coordinates,
the matrix $Q^{-1}-\delta$ does not contribute to the first square root
on the right-hand side of (\ref{dielectricBI}) and we get:
\beq
\sqrt{-\det\big[ P[G]\,\big]}\,=\,
{\rho^2+L^2\over R^2}\,\,
\sqrt{1\,+\,{(\rho^2+L^2)^2\over R^4}\,x'^{\,2}}\,\,.
\eeq
Moreover, by using the ansatz (\ref{Yansatz}) and the commutation
relations (\ref{Jcommutator}) we obtain that, for large $k$, the second
square root appearing in (\ref{dielectricBI}) can be written as:
\beq
{\rm Str}\Bigg[\sqrt{\det Q}\Bigg]\,\approx\,
{R^2\over \pi\alpha'}\,\,
{\rho^2\over \rho^2+L^2}\,\,
\sqrt{1\,+\,{(\rho^2+L^2)^2\over R^4}\,
{(k\pi\alpha')^2\over \rho^4}}\,\,.
\eeq

Using these results, the Born-Infeld part of the D3-brane action
in this large $k$ limit takes the form:
\beq
S_{BI}^{D3}\,=\,-{T_3\over \pi\alpha'}\,\,
\int\,d^3x\,d\rho\,
\rho^2\,
\sqrt{1\,+\,{(\rho^2+L^2)^2\over R^4}\,x'^{\,2}}\,\,
\sqrt{1\,+\,{(\rho^2+L^2)^2\over R^4}\,
{q^2\over \rho^4}}\,\,,
\label{microBI}
\eeq
where we have already used (\ref{q-k}) to write the result in terms of
$q$.  Due to the relation $T_3\,=\,4\pi^2\,\alpha'\,T_5$ between the
tensions of the D3- and D5-branes, one checks by inspection that the
right-hand side of (\ref{microBI}) coincides with the Born-Infeld term of
the D5-brane action (\ref{effe-action}). Notice also that the quantization
integer $k$ in (\ref{fluxquantization}) is identified with the number of
D3-branes.  Moreover, the Wess-Zumino term (\ref{dielectricWZ}) becomes:
\beq
S_{WZ}^{D3}\,=\,kT_3\,\int\,d^3x\,d\rho\,\,\,
{(\rho^2+L^2)^2\over R^4}\,x'\,\,.
\label{microWZ}
\eeq
The factor $k$ in (\ref{microWZ}) comes from the trace of the unit
$k\times k$ matrix. By comparing (\ref{microWZ}) with the Wess-Zumino
term of the  macroscopical action  (\ref{effe-action}) one readily
concludes that they  coincide because of the relation $4\pi
qT_5\,=\,kT_3$, which can be easily proved.

\section{Field theory analysis}	
\label{fieldtheory}

In this section we will analyze the configuration described above from the
point of view of the field theory at the defect which, from now on, we
shall assume that it is located at $x^3=0$. Recall that the defect arises
as a consequence of the impurity created on the D3-brane worldvolumes by
the D5-brane which intersects with them according to the array
(\ref{D3D5intersection}).  We are interested in analyzing, from the field
theory point of view, the configurations in which some fraction of the
D3-branes end on the D5-brane and recombine with it at the defect point
$x^3=0$, realizing in this way a (mixed Coulomb-) Higgs branch of the
defect theory.

The field theory dual to the D3-D5 intersection 
has been worked out by DeWolfe \textit{et al.} in ref. \cite{WFO}. The
theory, which includes ${\mathcal N}=4$ $SU(N)$ $SYM$ in 4d plus an
${\cal N}=4$ hypermultiplet confined to the defect, has an $SU(2)_H\times
SU(2)_V$ R-symmetry. The $SU(2)_H$ ($SU(2)_V$) symmetry corresponds to
the rotations in the 456 (789) directions of the array 
(\ref{D3D5intersection}). Written in terms of ${\mathcal N}=1$ SUSY, this
hypermultiplet gives rise to a chiral ($Q$) and an antichiral ($\bar{Q}$)
supermultiplet, which are both doublets under $SU(2)_H$ while being in
the fundamental representation of the gauge group. In addition, the 6
scalars of the bulk ${\mathcal N}=4$, which are in the adjoint of the
gauge group, naturally split in two sets, the first (which we will call
$\phi_H^I$) forming a vector of $SU(2)_H$ and the second,  which we denote
by $\phi_V^A$,  a vector of $SU(2)_V$. Thus, 
the bosonic content of the theory is:

\begin{center}
\begin{tabular}{|c|c|c|c|}
\hline 
Field & $SU(N)$ & $SU(2)_H$ & $SU(2)_V$\\
\hline
$A_{\mu}$ & adjoint & singlet & singlet\\
\hline
$\phi_H^I$ & adjoint & vector & singlet\\
\hline
$\phi_V^A$ & adjoint & singlet & vector\\
\hline
$q$ & fundamental & doublet & singlet\\
\hline
$\bar{q}$&fundamental& doublet & singlet\\
\hline
\end{tabular}
\label{table}
\end{center}
We will assume that only the fields $\phi_H$, $\phi_V$, $q$ and $\bar q$
are non-vanishing. The defect action for this theory has a potential term
which can be written as \cite{WFO}:
\begin{eqnarray}
S_{defect}&=&-\frac{1}{g^2}\int d^3x\,\Bigg[\,
 \bar{q}^m\,(\phi_V^A)^2\,q^m+
\frac{i}{2}\epsilon_{IJK}\bar{q}^m\sigma_{mn}^I\,[\phi_H^J,\phi_H^K]\,q^n
\,\Bigg]\,-\,\rc
&-&\frac{1}{g^2}\int d^3x\Bigg[\,
\bar{q}^m\sigma_{mn}^I\partial_3\,\phi_H^I\,q^n+
\frac{1}{2}\delta(x_3)(\bar{q}^m\sigma_{mn}^IT^aq^n\,)^2\,\Bigg]\,\,,
\label{actiondefect}
\end{eqnarray}
where the integration is performed over the $x^3=0$ three-dimensional
submanifold and $g$ is the Yang-Mills coupling constant. In the
supersymmetric configurations we are looking for the potential term must
vanish. Let us cancel the contribution of $\phi_V$ to the right-hand side 
of (\ref{actiondefect}) by requiring that: 
\beq
\phi_V\,q\,=\,0\,\,.
\eeq
We can insure this property by taking $q$ as:
\beq
q=\left(\begin{array}{c}0 \\ \vdots\\ 0\\ 
\alpha_1\\ \vdots \\ \alpha_k\end{array}\right)\,\,,
\label{qvev}
\eeq
and by demanding that  $\phi_V$ is of the form:
\beq
\phi_V\,=\,\pmatrix{A&0\cr
0&0}\,\,,
\eeq
where $A$ is an $(N-k)\times (N-k)$ traceless matrix. Moreover, we shall
take $\phi_V$, $q$ and $\bar q$ constant, which is enough to guarantee
that their kinetic energy vanishes. Notice that the scalars $\phi_V$
correspond to the directions 789 in the array (\ref{D3D5intersection}),
which are orthogonal to both the D3- and D5-brane. Having $\phi_V\not=0$
is equivalent to taking $|\vec z|=L\not=0$ in the approach of sections
\ref{fluxD3D5} and \ref{dielectric}, and it corresponds to a non-zero
value of the mass of the hypermultiplets (see the first term in the
defect action (\ref{actiondefect})).

Let us now consider the configurations of $\phi_H$ with vanishing energy.
First of all we will impose that $\phi_H$ is a matrix whose only
non-vanishing entries are in the lower $k\times k$ block. In this way the
mixing terms of $\phi_V$ and $\phi_H$ cancel. Moreover, assuming that
$\phi_H$ only depends on the coordinate $x^3$, the surviving terms in the
bulk action are \cite{WFO}:
\beq
S_{bulk}\,=\,-{1\over g^2}\,\,
\int\,d^4x\,{\rm Tr}\,\Bigg[\,
{1\over 2}\,\,(\partial_3\phi_H^I)^2\,-\,{1\over 4}\,\,
[\phi_H^I, \phi_H^J]^2\,\,\Bigg]\,\,,
\label{bulkaction}
\eeq
where the trace is taken over the color indices. It turns out that the
actions (\ref{actiondefect}) and (\ref{bulkaction}) can be combined in
such a way that their sum can be written as an integral over the
four-dimensional spacetime of the trace of
a square. In order to write this
expression, let  us  define the matrix
$\alpha^I\,=\,\alpha^{Ia}\,\,T^a$, where the $T^a$'s are the generators
of the gauge group and the $\alpha^{Ia}$'s are defined as the following
expression bilinear in $q$ and $\bar q$:
\beq
\alpha^{Ia}\,\equiv 
\bar{q}^m\,\sigma_{mn}^I\,T^a\,q^n\,\,.
\eeq
It is now straightforward to check that the sum of (\ref{actiondefect})
and (\ref{bulkaction}) can be put as:
\beq
S_{defect}\,+\,S_{bulk}\,=\,-{1\over 2g^2}\,\,
\int d^4x {\rm Tr}\,\Bigg[\,
\partial_3\phi_H^I\,+\,{i\over 2}\,
\epsilon_{IJK}\,[\,\phi_H^J, \phi_H^K\,]\,+\,
\alpha^I\,\delta(x^3)\,\,\Bigg]^2\,\,,
\label{actionsquare}
\eeq
where we have used the fact that
$\epsilon_{IJK}\,{\rm Tr}\,\Big(\,\partial_3\phi_H^I\,
[\,\phi_H^J, \phi_H^K\,]\,\Big)$ is a total derivative with respect to
$x^3$ and, thus, can be dropped if we assume that $\phi_H$ vanishes at 
$x^3=\pm\infty$. It is now clear from (\ref{actionsquare}) that we must
require the Nahm equations \cite{Neq}: 
\beq
\partial_3 \phi_H^I\,+\,{i\over 2}\,\epsilon_{IJK}
[\,\phi_H^J, \phi_H^K\,]\,+\,\alpha^I\,\delta(x^3)
\,=\,0\,\,.
\label{Nahn}
\eeq
(For a nice review of the Nahm construction in string theory see 
\cite{Tong}). 

Notice that when $\alpha^I$ vanishes, eq. (\ref{Nahn}) admits the trivial
solution $\phi_H=0$. On the contrary, if the fundamentals $q$ and $\bar
q$ acquire a non-vanishing vacuum expectation value
as in (\ref{qvev}),  $\alpha^I$  is
generically non-zero and the solution of (\ref{Nahn}) must be
non-trivial. Actually, it is clear from (\ref{Nahn}) that  in this case
$\phi_H$ must blow up at $x^3=0$, which shows how a non-vanishing vacuum
expectation value of the fundamentals acts as a source for the brane
recombination in the Higgs branch of the theory. Let us check these facts
more explicitly by solving (\ref{Nahn}) for $x^3\not=0$, where the
$\delta$-function term is zero. We shall adopt the ansatz:
\beq
 \phi_H^I(x)\,=\,f(x)\,\phi_0^I\,\,,
 \eeq
where $x$ stands for $x^3$ and $\phi_0^I$ are constant 
matrices.  The differential
equation (\ref{Nahn}) reduces to: 
\beq
{f'\over f^2}\,\,\phi_0^I\,+\,{i\over 2}\,
\epsilon_{IJK}\,[\,\phi_0^J\,,\phi_0^K\,]\,=\,0\,\,,
\label{soucelessNahm}
\eeq
where the prime denotes derivative with respect to $x$. 
We shall solve this equation by first putting:
\beq
\phi_0^I\,=\,{1\over \sqrt{C_2(k)}}\,
\pmatrix{0&0\cr 0&J^I}\,\,,
\eeq
where the $J^I$ are matrices in the $k$-dimensional
irreducible representation of the $SU(2)$ algebra, which
satisfy the commutation relations (\ref{Jcommutator}), 
and we have normalized the $\phi_0^I$'s such that 
$\phi_0^I\phi_0^I$ is the unit matrix in the $k\times k$ block. 
By using this representation of the $\phi_0^I$'s,  eq.
(\ref{soucelessNahm}) reduces to:
\beq
{f'\over f^2}\,=\,{2\over \sqrt{C_2(k)}}\,\,,
\eeq
which can be immediately integrated, namely:
\beq
f\,=\,-{\sqrt{C_2(k)}\over 2 x}\,\,.
\eeq
For large $k$, the quadratic Casimir $C_2(k)$ behaves as $k^2$ and this
equation reduces to:
\beq
f\,=\,-{k\over 2 x}\,\,.
\label{Nahnsolution}
\eeq
Let us now take  into account the standard relation between coordinates
$X_H^I$ and  scalar fields $\phi^I_H$, namely:
\beq
X_H^I\,=\,2\pi\alpha'\,\phi^I_H\,\,,
\eeq
and the fact that $\rho^2\,\equiv\,X_H^I\,X_H^I$. Using these facts we
immediately get the following relation between $\rho$ and $f$:
\beq
\rho\,=\,2\pi\alpha' f\,\,,
\eeq
and the solution (\ref{Nahnsolution})
 of the Nahm equation can be written as:
\beq
\rho\,=\,-{\pi k\alpha'\over x}\,\,,
\eeq
which, if we take into account the quantization condition (\ref{q-k}), 
is just our embedding (\ref{bending}) for $x_0=0$. 
As expected, $\rho$ blows up at
$x=0$, while its dependence for $x\not=0$ gives rise to the same bending
as in the brane approach. Notice also that, in this field theory
perspective, the integer $k$ is the rank of the gauge theory subgroup in
which the Higgs branch of the theory is realized, which corresponds to
the number of D3-branes that recombine into a D5-brane.

\section{Fluctuations}
\label{fluctuations}
Let us now analyze the small fluctuations around the static embedding of
the D5-brane probe described in section. \ref{fluxD3D5}.  For simplicity
we will restrict ourselves to study the fluctuations of the scalars
transverse to both the D3- and D5-branes (\ie\ those along the directions
789 in the array (\ref{D3D5intersection})). It can be shown that, at
quadratic order, these fluctuations do not couple to those corresponding
to the worldvolume gauge field and the scalar $x^3$. In the unperturbed
configuration the distance $|\vec z|$ between the two types of branes is
constant and equal to $L$ and, without loss of generality, we can assume
that the branes are separated along the $z^1$ direction. Accordingly, let
us consider a fluctuation of the type:
\beq
z^1\,=\,L\,+\,\chi^1\,\,,
\qquad z^2\,=\,\chi^2\,\,,\qquad
\qquad z^3\,=\,\chi^3\,\,,
\eeq
where the $\chi^m$ are small. 
The induced metric for this perturbed configuration
can be decomposed as:
\beq
g\,=\,{\cal G}\,+\,g^{(f)}\,\,,
\eeq
where ${\cal G}$ is the metric written in (\ref{ind-met-flux}) and
$g^{(f)}$ is the  part of $g$ that depends
on the derivatives of the fluctuations, namely:
\beq
g^{(f)}_{ab}\,=\,
{R^2\over \rho^2+L^2}\,\partial_a \chi^m\,\partial_b \chi^m\,\,.
\eeq
The Born-Infeld  determinant in the action (\ref{DBI-D5}) can be written
as:
\beq
\sqrt{-\det(g+ F)}\,=\,\sqrt{-\det \big(\,{\cal G}\,+\,{\cal
F}\,\big)}\,
\sqrt{\det\,(1+M)}\,\,,
\label{detX}
\eeq
where ${\cal F}$ is the worldvolume gauge field (\ref{wvflux}) and
the matrix $M$ is given by:
\beq
M\,\equiv\,\bigg(\,{\cal G}\,+\,{\cal  F}\,\bigg)^{-1}\,\,
\,g^{(f)}\,\,.
\label{matrixM}
\eeq
To evaluate  the right-hand side of eq. (\ref{detX}), we shall use
the expansion:
\beq
\sqrt{\det\,(1+M)}\,=\,1\,+\,{1\over 2}\,\tr M\,+\,o(M^2)\,\,.
\label{expansion}
\eeq
The prefactor multiplying this expansion in (\ref{detX}) is:
\beq
\sqrt{-\det \big(\,{\cal G}\,+\,{\cal
F}\,\big)}\,=\,\rho^2\sqrt{\tilde g}\,\bigg(\,
1\,+\,{q^2\over R^4}\,{(\rho^2+L^2)^2\over \rho^4}\,\bigg)\,\,,
\eeq
where $\tilde g$ is the determinant of the round metric for the unit 
two-sphere. Moreover, let us separate the symmetric and antisymmetric
part in the inverse matrix  appearing in the  expression of
 $M$ (eq. (\ref{matrixM})):
\beq
\bigg(\,{\cal G}\,+\,{\cal  F}\,\bigg)^{-1}\,=\,
\hat{\cal G}^{-1}\,+\,{\cal J}\,\,,
\eeq
where:
\beq
\hat{\cal G}^{-1}\,=\,{1\over ({\cal G}+F)_S}\,\,,\qquad
{\cal J}\,=\,{1\over ({\cal G}+F)_A}\,\,.
\eeq
Notice that $\hat{\cal G}$ is just the open string metric.
After a straightforward calculation one can verify that $\hat{\cal G}$
can be written as:
\beq
\hat{\cal G}_{ab}d\xi^a\,d\xi^b\,=\,{\rho^2+L^2\over R^2}\,
dx_{1,2}^2\,+\,{R^2\over \rho^2+L^2}\,\bigg(\,
1\,+\,{q^2\over R^4}\,{(\rho^2+L^2)^2\over \rho^4}\,\bigg)\,
\bigg(\,d\rho^2\,+\,\rho^2\,d\Omega_2^2\,\bigg)\,\,.
\label{effect-metric}
\eeq
Moreover, the antisymmetric matrix ${\cal J}$ has only non-vanishing
values when its two-indices are on the two-sphere. Actually, if $\theta$
and $\varphi$ are the standard polar coordinates on $S^2$, we have:
\beq
{\cal J}^{\theta\varphi}\,=\,-{\cal J}^{\varphi\theta}\,=\,
-{1\over \sqrt{\tilde g}}\,\,
{q\over q^2\,+\,{R^4\rho^4\over (\rho^2+L^2)^2}}\,\,.
\eeq
Notice that the antisymmetric matrix  ${\cal J}$ does not contribute to 
$\tr M$ since it is contracted with $g^{(f)}$, which is symmetric. The
final result for  $\tr M$ is:
\beq
\tr M\,=\,{R^2\over \rho^2+L^2}\,\,\hat{\cal 
G}^{ab}\,\partial_a\chi^m\,\partial_b\chi^m\,\,.
\eeq
By using this result we get that the  total lagrangian density for the
$\chi$ fluctuations  is given by:
\beq
{\cal L}\,=\,-\rho^2\,{\sqrt{\tilde g}\over 2}\,
{R^2\over  \rho^2+~L^2}\,\bigg(\,1\,+\,{q^2\over R^4}\,
{(\rho^2+L^2)^2\over \rho^4}\,\bigg)\,
\hat{\cal G}^{ab}\,\partial_a\chi^m\,\partial_b\chi^m\,\,.
\label{fluct-lag}
\eeq
It is clear from (\ref{fluct-lag}) that the open string metric 
$\hat{\cal G}$ governs the dynamics of the fluctuations. For this reason,
it is interesting to look at $\hat{\cal G}$ closely and, in particular to
compare it with the induced metric ${\cal G}$ of eq.
(\ref{ind-met-flux}). Notice that ${\cal G}$ and $\hat{\cal G}$ only
differ in the term corresponding to the two-sphere. Actually, 
the metric $\hat{\cal G}$ in the UV ($\rho\to\infty$) becomes 
$AdS_4(R_{eff})\times S^2 (R_{eff})$, where $R_{eff}$ is the effective 
radius of eq. (\ref{Reff}) (compare this result with
(\ref{UVind-met-flux})).  If the separation distance $L$ is zero 
${\cal G}$ and $\hat{\cal G}$ retain this $AdS\times S$ form for all
values of $\rho$. However, when $L\not= 0$ the IR behaviour of these
metrics changes drastically. Actually, it is clear from
(\ref{ind-met-flux}) that the $S^2$ factor in ${\cal G}$ collapses at
$\rho=0$.  On the contrary, when $q\not=0$, the terms with $q$ in 
$\hat{\cal G}$ dominate over the others in the IR  and 
the open string metric takes the form:
\beq
\hat{\cal G}_{ab}d\xi^a\,d\xi^b\,\approx\,{L^2\over R^2}\,\,
\Bigg[\,dx_{1,2}^2\,+\,q^2\Big(\,
{d\rho^2\over \rho^4}\,+\, {1\over \rho^2}\,d\Omega_2^2\,\Big)
\,\Bigg]\,\,,
\qquad\qquad(\rho\approx 0)\,\,.
\label{IRmetric}
\eeq
Notice that now $d\Omega_2^2$ is multiplied by a factor that diverges 
for $\rho\approx 0$ in (\ref{IRmetric}). Actually, by
changing variables from $\rho$ 
to  $u=q/\rho$, this metric can be written as:
\beq
{L^2\over R^2}\,\,\Big[\,dx_{1,2}^2\,+\, du^2\,+\,u^2\,
d\Omega_2^2\,\Big]\,\,,
\label{IRMinkowski}
\eeq
which is nothing but the six dimensional Minkowski space.

The equation of motion for the transverse scalars $\chi$ derived from 
(\ref{fluct-lag}) is:
\beq
\partial_a\Bigg[\sqrt{\tilde g}\,{\rho^2\over \rho^2+L^2}\,
\bigg(\,1\,+\,{q^2\over R^4}\,{(\rho^2+L^2)^2\over \rho^4}\,\bigg)\,
\hat{\cal G}^{ab}\,\partial_b\chi\,\Bigg]\,=\,0\,\,.
\eeq
By using the explicit form of the effective metric $\hat{\cal G}^{ab}$ (eq.
(\ref{effect-metric})), we can write this equation as:
\beq
\bigg[ {R^4\rho^2\over (\rho^2+L^2)^2}\,+\,{q^2\over \rho^2}\,\bigg]\,\,
\partial^{\mu}\partial_{\mu}\,\chi\,+\,
\partial_{\rho}\,\Big(\rho^2\partial_\rho 
\chi\Big)\,+\,\nabla^i\nabla_i\,\chi\,=\,0\,\,,
\label{eom-fluc}
\eeq
where the indices $\mu$ correspond to the $2+1$ Minkowski directions
and the $i$'s are those of the two-sphere. Let us next separate variables
and  write the scalars $\chi$  in terms of the spherical harmonics on
the two-sphere and plane waves in the Minkowski coordinates:
\begin{equation}
\chi=e^{ikx}\,Y^l(S^2)\,\xi(\rho)\,\,,
\end{equation}
where the product $kx$ is performed with the flat Minkowski metric and
$l$ denotes the angular momentum on the $S^2$. The mass of the meson is 
defined as $M^2=-k^2$.  By using this ansatz, 
the equation of motion (\ref{eom-fluc}) reduces to:
\beq
\partial_{\rho}\,\Big(\rho^2\partial_{\rho}\xi\Big)\,+\,\Bigg\{\,\bigg[
{R^4\rho^2\over (\rho^2+L^2)^2}\,+\,{q^2\over
    \rho^2}\,\bigg]\,M^2-l(l+1)\,\Bigg\}\xi\,=0\, .
\label{transvxieq}
\eeq

Let us analyze the solutions of (\ref{transvxieq}) when the distance
$L\not=0$. In general, the requirement of having a regular normalizable
solution for the fluctuation $\xi$ determines the allowed values of the
mass $M$. Let us now look at this regularity condition at the UV. For
$\rho\to\infty$ one can easily show that there are two independent
solutions of (\ref{transvxieq}) which behave as $\rho^{l}$ or
$\rho^{-l-1}$. Clearly, the admissible fluctuations are the ones
decreasing as $\xi\sim\rho^{-l-1}$ for large $\rho$. For $q=0$ (\ie\
without flux) the fluctuations also behave as $\rho^{\gamma}$, with 
$\gamma=l, -l-1$ when $\rho\approx 0$ and, thus,  we should impose that
$\xi\sim\rho^{l}$ at the IR. The matching of the $\rho\approx 0$
behaviour with that for $\rho\to\infty$ is only possible for a discrete
set of values of the mass $M$. The corresponding discrete spectrum was
found analytically in \cite{open} and has a mass gap proportional to
$L/R^2$. However, when
$q\not=0$ the 
$\rho\approx 0$ behaviour of the solutions of (\ref{transvxieq}) changes
drastically. Indeed, when $\rho$ is small and $q$ does not vanish
eq. (\ref{transvxieq}) reduces to:
\beq
\partial_{\rho}\,\Big(\rho^2\partial_{\rho}\xi\Big)\,+\,
\Bigg[\,{q^2\,M^2\over\rho^2}\,-\,l(l+1)\,\Bigg]\xi\,=0\,\,,
\qquad\qquad(\rho\approx 0)\,\,.
\label{IRfluc}
\eeq
Eq.  (\ref{IRfluc}) can be solved in terms of Bessel functions,
namely:
\beq
\xi\,=\,{1\over \sqrt{\rho}}\,\,
J_{\pm (l+{1\over 2})}\,\Bigg({qM\over \rho}\Bigg),\,
\qquad\qquad(\rho\approx 0)\,\,.
\label{Bessel}
\eeq
Near $\rho\approx 0$ the Bessel function (\ref{Bessel}) behaves as:
\beq
\xi\,\approx\,e^{\pm i{qM\over \rho}}\,\,,
\qquad\qquad(\rho\approx 0)\,\,,
\label{IRfluct-behaviour}
\eeq
\ie\ it oscillates infinitely as we approach $\rho=0$. Notice that this
gives rise to a continuous gapless spectrum for M. Actually, one can
understand this result by rewriting the function (\ref{Bessel}) in terms
of the coordinate $x^3$ by using (\ref{bending}). Indeed, $\rho\approx 0$
corresponds to large
$|x^3|$ and $\xi(x^3)$ can be written in this limit as 
a simple plane wave:
\beq
\xi\,\approx\,e^{\pm iMx^3}\,\,,\qquad\qquad(\,|x^3|\to\infty\,)\,\,.
\eeq
Notice that this behaviour is consistent with the fact that the IR metric
for $q\not=0$ approaches the Minkowski metric (\ref{IRMinkowski}) and the
fluctuation spreads out of the defect locus $x^3=0$. We  have checked
that this fact is generic by analyzing the full set of fluctuations of
the D3-D5 system. These fluctuations are coupled, but it turns out that
they can be decoupled by using the same techniques as those employed in
ref. \cite{open} for the $q=0$ case. One can show that these decoupled
functions have the same qualitative behaviour as in
(\ref{IRfluct-behaviour}), which implies that the spectrum is continuous
for $q\not=0$ \cite{progress}.

\section{Concluding Remarks}
\label{conclusions}
In the field theory dual, the addition of a brane probe to a given
background corresponds to the insertion of an impurity, which is
generically located at a defect in the gauge theory directions. This
defect hosts new open string degrees of freedom  (hypermultiplets) which
interact non-trivially with the original bulk fields. The AdS/CFT
correspondence can be used to obtain the mass spectra of the mesonic
operators, which are bilinear in the hypermultiplet fields, by studying
the fluctuations of the flavor branes \cite{KKW}. As reviewed in
\cite{MPLAreview},  when the hypermultiplets
have a non-vanishing mass, one gets a well-defined discrete spectrum with
a non-zero mass gap.

In this paper we have explored the effect of giving a vacuum expectation
value to some components of the hypermultiplets. In string theory, such a
Higgs branch is realized by recombining color and flavor branes in a
non-trivial way. From the point of view of the worldvolume theory of the
flavor brane, this recombination is realized by adding a suitable flux of
the worldvolume gauge field, such that some units of charge of the color
brane are dissolved in the worldvolume of the flavor brane. Then,
supersymmetry requires that the flavor brane must be bent appropriately,
as in eq. (\ref{bending}).

A natural question, which we addressed in this paper for the defect
theory dual to the D3-D5 intersection, is how the meson spectrum is
affected when one moves to the Higgs branch. We have shown above that the
mesonic mass spectrum of the defect theory is drastically changed (it
becomes continuous and gapless). The reason behind this behaviour is the
fact that the color and flavor branes are connected to each other, which
delocalizes the fluctuations in the direction orthogonal to the defect.
Actually, this behaviour is generic of any intersection dual to a defect
theory with codimension greater than zero \cite{progress}. On the
contrary, this does not happen in the cases in which the flavor brane
fills completely the gauge theory directions, such as the D3-D7
intersection with a worldvolume instanton studied in ref. \cite{Higgs}.
We expect to report on these results in ref. \cite{progress}.

\section*{Acknowledgments}

We are grateful to J. D. Edelstein,  M. Kruczenski, C. N\'u\~nez and to
the participants of the Ringberg Castle workshop 
``QCD and String Theory" (where a prelimiray version of this work was
presented) for discussions. The  work of DA and AVR was
supported in part by MCyT, FEDER and Xunta de Galicia under grant
FPA2005-00188 and by  the EC Commission under  
grants HPRN-CT-2002-00325 and MRTN-CT-2004-005104. DRG is grateful to
Universidad Autonoma de Madrid, Universidad de Granada and Universidade
de Santiago de Compostela for their hospitality and support during
various stages of this work. His work has been also partially supported
by CICYT grant BFM2003-00313 (Spain) and by the EC Comission FP6
program MRTN-CT-2004-005104 in which he is associated to UAM.

\end{document}